\documentclass[numberedappendix,natbib209]{emulateapj}

\usepackage{epsfig}
\usepackage{longtable}
\usepackage{natbib}

\begin{document}

\title{A Strongly-Lensed Massive Ultra-Compact Quiescent Galaxy at \lowercase{z} $\sim$ 2.4 in the COSMOS/UltraVISTA Field\altaffilmark{1}}
\author{Adam Muzzin\altaffilmark{2}, Ivo Labb\'{e}\altaffilmark{2}, Marijn Franx\altaffilmark{2}, Pieter van Dokkum\altaffilmark{3}, J. Holt\altaffilmark{2}, Daniel Szomoru\altaffilmark{2}, Jesse van de Sande\altaffilmark{2}, Gabriel Brammer\altaffilmark{4}, Danilo Marchesini\altaffilmark{5}, Mauro Stefanon\altaffilmark{6}, F. Buitrago\altaffilmark{7}, K. I. Caputi\altaffilmark{8}, James Dunlop\altaffilmark{7}, J. P. U. Fynbo\altaffilmark{9}, Olivier Le F\'{e}vre\altaffilmark{10}, Henry J. McCracken\altaffilmark{11}, Bo Milvang-Jensen\altaffilmark{9}}

\altaffiltext{1}{Based on data products from observations made with ESO Telescopes at the La Silla Paranal Observatory under ESO programme ID 179.A-2005 and on data products produced by TERAPIX and the Cambridge Astronomy Survey Unit on behalf of the UltraVISTA consortium.}

\altaffiltext{2}{Leiden Observatory, Leiden University, PO Box 9513,
  2300 RA Leiden, The Netherlands}
\altaffiltext{3}{Department of Astronomy, Yale
  University, New Haven, CT, 06520-8101} 
\altaffiltext{4}{European Southern Observatory, Alonso de C\'{o}rdova 3107, Casilla 19001, Vitacura, Santiago, Chile} 
\altaffiltext{5}{Department of Physics and Astronomy, Tufts University, Medford, MA 06520, USA}
\altaffiltext{6}{Obseratori Astron\`{o}mic de la Universitat de Val\`{e}ncia, 46980 Paterna, Val\`{e}ncia, Spain}
\altaffiltext{7}{SUPA, Institute for Astronomy, University of Edinburgh, Royal Observatory, Edinburgh EH9 3HJL, UK}
\altaffiltext{8}{Kapteyn Astronomical Institute, University of Groningen, P.O. Box 800, 9700
AV Groningen, The Netherlands.}
\altaffiltext{9}{Dark Cosmology Centre, Niels Bohr Institute, University of Copenhagen, Juliane Maries Vej 30, 2100 Copenhagen, Denmark}
\altaffiltext{10}{Laboratoire d'Astrophysique de Marseille, CNRS and Aix-Marseille Universit\'{e}, 38 rue Fr\'{e}d\'{e}ric Joliot-Curie, 13388 Marseille, Cedex 13, France}
\altaffiltext{11}{Institut d'Astrophysique de Paris, UMR7095 CNRS, Universit\'{e} Pierre et Marie Curie, 98 bis Boulevard Arago, 75014 Paris, France}

\begin{abstract}
We report the discovery of a massive ultra-compact quiescent galaxy that has been strongly-lensed into multiple images by a foreground galaxy at $z = 0.960$.  This system was serendipitously discovered as a set of extremely K$_{s}$-bright high-redshift galaxies with red J - K$_{s}$ colors using new data from the UltraVISTA YJHK$_{s}$ near-infrared survey.  The system was also previously identified as an optically-faint lens/source system using the COSMOS ACS imaging by Faure et al. (2008, 2011).   Photometric redshifts for the three brightest images of the source galaxy determined from twenty-seven band photometry place the source at $z =$ 2.4 $\pm$ 0.1.  We provide an updated lens model for the system which is a good fit to the positions and morphologies of the galaxies in the ACS image.  The lens model implies that the magnification of the three brightest images is a factor of 4 - 5.   We use the lens model, combined with the K$_{s}$-band image to constrain the size and Sersic profile of the galaxy.  The best-fit model is an ultra-compact galaxy (R$_{e}$ = 0.64$^{+0.08}_{-0.18}$ kpc, lensing-corrected), with a Sersic profile that is intermediate between a disk and bulge profile (n = 2.2$^{+2.3}_{-0.9}$), albeit with considerable uncertainties on the Sersic profile.   
We present aperture photometry for the source galaxy images which have been corrected for flux contamination from the central lens.  The best-fit stellar population model is a massive galaxy (Log(M$_{star}$/M$_{\odot}$) = 10.8$^{+0.1}_{-0.1}$, lensing-corrected) with an age of 1.0$^{+1.0}_{-0.4}$ Gyr, moderate dust extinction (A$_{v}$ = 0.8$^{+0.5}_{-0.6}$), and a low specific star formation rate (Log(SSFR) $<$ -11.0 yr$^{-1}$).   This is typical of massive ``red-and-dead" galaxies at this redshift and confirms that this source is the first bona fide strongly-lensed massive ultra-compact quiescent galaxy to be discovered.  We conclude with a discussion of the prospects of finding a larger sample of these galaxies.
%
\end{abstract}
\keywords{infrared: galaxies -- galaxies: evolution -- galaxies: high-redshift -- galaxies: structure  -- galaxies: fundamental parameters}

\section{Introduction}
\begin{figure*}
\plotone{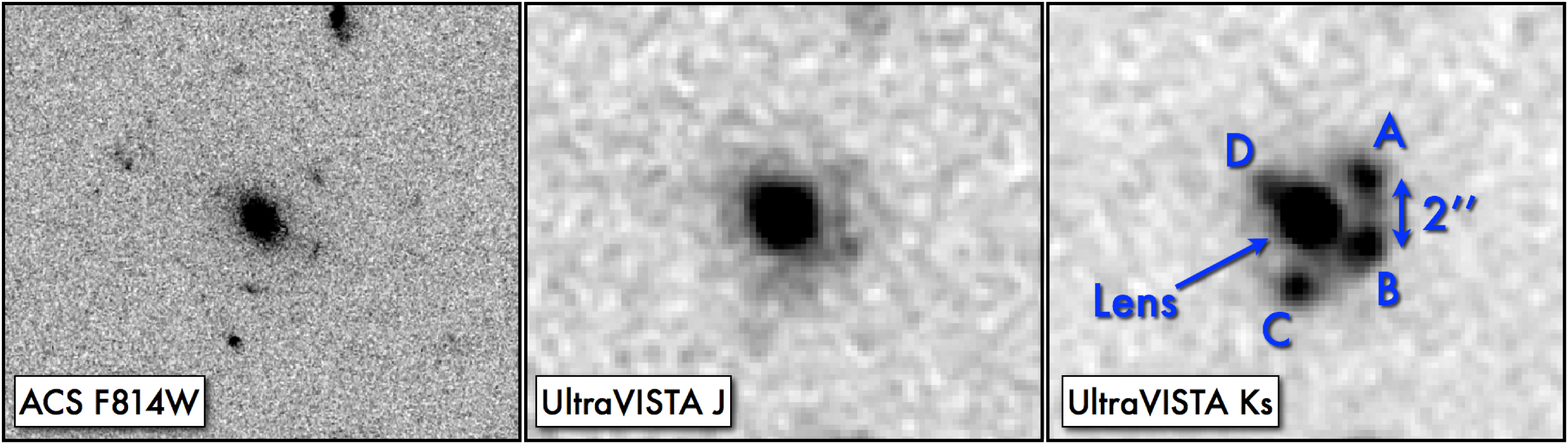}
\caption{\footnotesize Postage stamp optical/NIR images of the lens/source system with a high-contrast stretch.  The field-of-view in each image is 12$^{\prime\prime}$ $\times$ 10$^{\prime\prime}$.  The lensed galaxy is faint in the observed optical, and has an extremely red J - K$_{s}$ color.  The separation between the multiple images is $\sim$ 2$^{\prime\prime}$.}
\end{figure*}
The brightening and magnifying effects from strong gravitational lensing are a powerful tool that permit us to study high-redshift galaxies with better signal-to-noise (S/N) and higher spatial resolution than is normally possible with current instrumentation.  In the last decade significant successes in finding samples of lensed galaxies have come from surveys that have searched behind clusters \citep[e.g.,][]{Bayliss2011b} and luminous red galaxies \citep[e.g.,][]{Bolton2008,Faure2008, Cooray2011}.  These samples are now giving us insights into the evolution of typical galaxies at 1 $< z <$ 5 which are most frequently low-mass blue star-forming galaxies \citep[e.g.,][]{Rigby2011, Wuyts2012, Sharon2012, Brammer2012}.    With the advent of the Herschel and South Pole Telescope submillimeter observatories we have also gained access to a population of strongly-lensed submillimeter galaxies which are readily identifiable in wide-field surveys \citep[e.g.,][]{Negrello2010,Vieira2010} or behind clusters \citep[e.g.,][]{Egami2010,Gladders2012}.  These lensed submillimeter galaxies are providing intriguing results about how dusty and strongly star forming galaxies evolve at high redshift \citep[e.g.,][]{Swinbank2010,Swinbank2011}.
\newline\indent
What has remained elusive are strongly-lensed examples of the proverbial ``red-and-dead" massive galaxies at $z \sim$ 2.  These galaxies have generated significant interest in the last few years after it was discovered that they are smaller in size than their similar-mass counterparts in the local universe by a factor of $\sim$ 5 \citep[e.g.,][and numerous others]{Daddi2005, Trujillo2006, vandokkum2008, Buitrago2008, vandokkum2010, Szomoru2010, Bruce2012}.  Understanding how they grow so substantially in size without adding significant stellar mass is currently one of the major open questions in galaxy evolution theory.  
\newline\indent
Finding gravitationally-lensed examples of these sources could prove to be valuable for understanding their evolution.  The magnifying effects from lensing will allow us to potentially resolve the central regions of these compact sources and understand what their central stellar density profiles are.  The brightening effects from lensing will also allow us to obtain higher S/N spectroscopy of these systems.  Currently, even the brightest of these systems are extremely faint and determining quantities such as velocity dispersions and dynamical masses requires substantial integration times even on the world's largest telescopes \citep[e.g.,][]{vandokkum2009,Cappellari2009,vandesande2011,Toft2012}.
\newline\indent
Unfortunately, these galaxies are much more challenging to detect as lensed sources and until now none have been securely identified.  The challenge is that while they comprise $\sim$ 50\% of the population of massive galaxies (LogM$_{star}$/M$_{\odot}$ $>$ 11.0) at $z \sim$ 2 \citep[e.g.,][]{Kriek2008,Brammer2011}, they are still quite rare ($\sim$ 1 every 10 arcminute$^{2}$), making it unlikely to find one in a favorable alignment with a foreground lensing structure.  Furthermore, they are extremely faint in the observed optical bands and also have very red J - K$_{s}$ NIR colors \citep[e.g.,][]{Franx2003, Kriek2008} making both deep $and$ wide-field NIR imaging a requirement for their detection.
\newline\indent
In this paper we report the discovery of the first example of a massive ultra-compact quiescent galaxy that has been strongly lensed by a foreground galaxy.  The galaxy was serendipitously discovered as a set of bright and extremely red high-redshift galaxies using new data from the 1.8 deg$^2$ UltraVISTA YJHK$_{s}$ near-infrared (NIR) survey \citep[see][]{McCracken2012}.  The lens/source system was previously identified as an optically faint quadruply-lensed galaxy in the COSMOS field using ACS imaging \citep{Faure2008,Faure2011}; however, with the addition of the UltraVISTA NIR data it is now clear that the source is a massive quiescent galaxy.  
\newline\indent
This paper is organized as follows, in $\S$ 2 we discuss how the lens/source system was discovered. In $\S$ 3 we present an updated lens model for the system and use this to constrain the size and Sersic profile of the source galaxy.  In $\S$ 4 we present updated twenty-seven band photometry for the source galaxies that has been deblended from contamination from the central source.  In $\S$ 5 we use this photometry to determine an accurate photometric redshift as well as stellar population parameters for the galaxy.  We conclude in $\S$ 6 with a summary and a discussion and prospects of finding more strongly-lensed massive ultra-compact quiescent galaxies in future deep, wide-field NIR surveys.  Throughout this paper we assume a $\Omega_{\Lambda}$ = 0.7, $\Omega_{m}$ = 0.3, and H$_{0}$ = 70 km s$^{-1}$ Mpc$^{-1}$ cosmology.  All magnitudes are in the AB system.
\section{Identification of the Lens/Source System}
The lens/source system was serendipitously discovered using photometry from a K$_{s}$-selected catalog of the COSMOS/UltraVISTA field.  The catalog contains photometry in twenty-seven photometric bands including the publically-available seven optical broad band (u$^{*}$g$^{+}$r$^{+}$i$^{+}$z$^{+}$B$_{j}$V$_{j}$) and twelve optical medium band (IA427 -- IA827) imaging of the COSMOS field from \cite{Capak2007}.  It also includes the public-release YJHK$_{s}$ NIR imaging of the COSMOS field from the UltraVISTA survey \citep{McCracken2012}, as well as the public four-channel IRAC imaging is from the S-COSMOS survey \citep{Sanders2007}.  Object detection for the catalog was performed in the K$_{s}$-band, which has a 5$\sigma$ depth in a 2$^{\prime\prime}$ aperture of K$_{s}$ = 23.9 AB \citep{McCracken2012}.  Photometry was measured using the Sextractor package \citep{Bertin1996} in 2.1$^{\prime\prime}$ apertures using PSF-matched images in all bands.  Full details of the COSMOS/UltraVISTA catalog will be presented in a future paper (A.~Muzzin et al., in preparation).
\newline\indent
The system (R.A., 10:00:50.55, Decl, +02:49:01, J2000) was identified as a strong-lensing system via eye-examination of the ACS F814W \citep{Koekemoer2007} and UltraVISTA K$_{s}$-band \citep{McCracken2012} images.  This process was performed as a quality check of the catalog for a subsample of galaxies that were were bright (K$_{s}$ $<$ 21.0 AB), red (J - K$_{s}$ $>$ 1.5 AB) and had photometric redshifts ($z_{photo}$) $>$ 2.  In Figure 1 we plot postage stamp images of the system in the ACS F814W, and UltraVISTA J and K$_{s}$ bands with a high-contrast stretch.  Sources A and C (see labels in Figure 1) were the sources in the catalog that matched the selection criteria.  Sources B and D lie along the major axis of the lens galaxy and hence are more blended with the lens and were not identified as unique sources by SExtractor. The extraordinary brightness of sources A and C given their $z_{photo}$, combined with the distinct cross-shape pattern of four sources around the brighter central lens (similar to quasar strong lenses) made it clear that the system was not just a very bright set of high-redshift galaxies, but in fact was a candidate strong-lensing system.  
\newline\indent
A literature search for known lens/source systems in the COSMOS field showed that this system had already been reported by \cite{Faure2008} and \cite{Faure2011}.  \cite{Faure2008,Faure2011} designated the system as ``COSMOS 0050+4901" and also noted that it is a clear quadruple-lens in the ACS images.  
\cite{Faure2011} report a spectroscopic redshift for the lens at $z_{spec}$ = 0.960.  No spectroscopic redshift was determined for the source galaxies which are extremely faint in the observed optical bands; however, \cite{Faure2008} estimate the redshift of the source as $z_{photo}$ = 3.34.  This does not agree well with our photometric redshift of the source ($\S$ 5.1); however, the source galaxies are extremely faint in the observed optical and their SEDs have no notable features there (see $\S$ 5.2) so it is unsurprising that a highly uncertain photometric redshift was estimated without high-quality NIR data such as that from UltraVISTA.
\newline\indent
The analysis presented in \cite{Faure2008,Faure2011} focuses on the properties of the lens galaxy.  In this paper we focus on the properties of the source galaxy which is currently the only-known candidate for a strongly-lensed compact quiescent galaxy.  
\begin{figure*}
\plotone{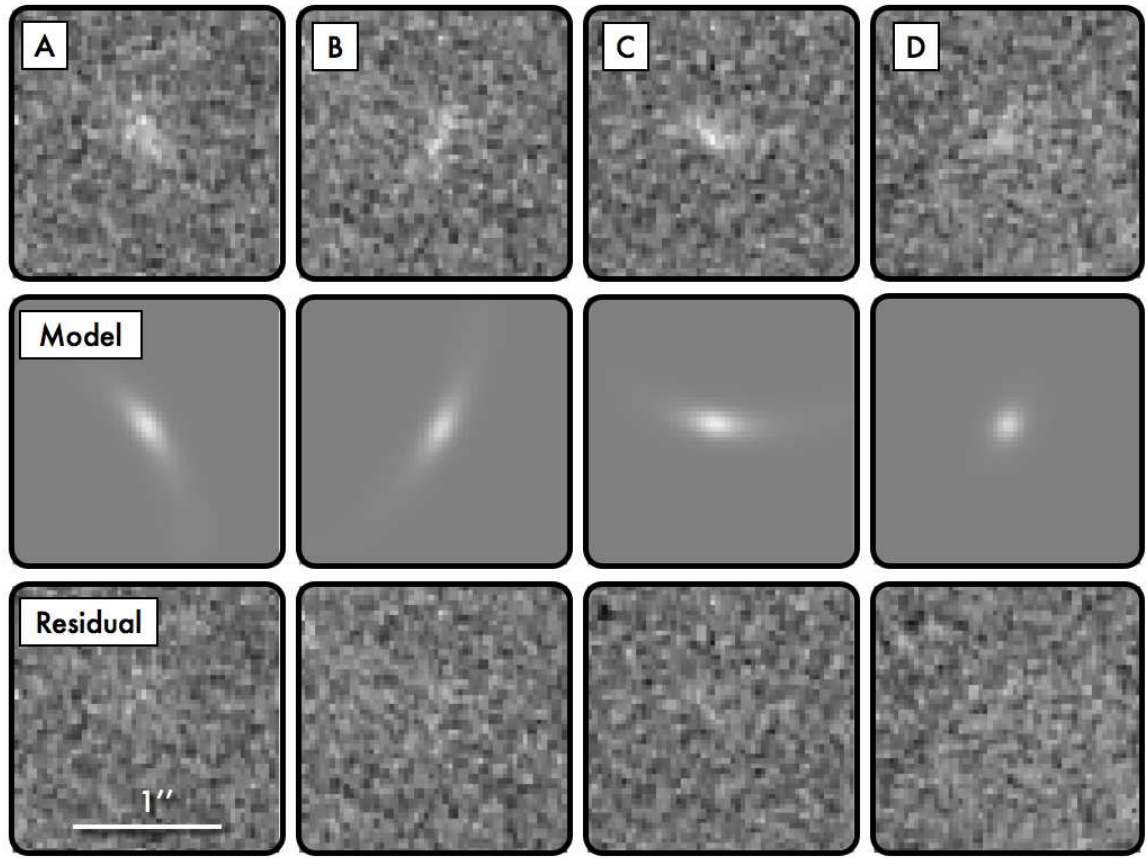}
\caption{\footnotesize Top Panels: ACS F814W postage stamps of the four images of the source galaxy.  Middle Panels: The best-fit ray-traced models for the source galaxy. Bottom Panels: Residuals of subtracting the model for the lens.  The best-fit model simultaneously reproduces the relative brightness and positions of the source, as well as its morphology and size.}
\end{figure*}
\section{Lens Model}
\indent
In order to measure the structural parameters of the source galaxy, we construct a model of the lens using the position and brightness of each of the four images of the source.  We assume the lens galaxy is embedded in a dark matter halo with an elliptical isothermal sphere profile with an additional core component.  The lens model has six free parameters: the Einstein radius (R$_{Ein}$), core radius (R$_{core}$), ellipticity ($e$), position angle (PA), and X and Y position of the lensed galaxy in the source plane (X$_{s}$, Y$_{s}$).  
\newline\indent
In order to constrain this large parameter space, we assumed a simple Sersic profile for the source galaxy with an effective radius (R$_{e}$) of 1.0 kpc, and a Sersic index ($n$) of 2.5.  We then generated a grid of models spanning the six-dimensional parameter space with fine sampling and used ray tracing to construct mock images at ACS resolution of the four images of the lensed galaxy.  The residuals from a subtraction of the ACS image and the mock images were then evaluated using $\chi^2$ to determine the best-fit lens model parameters.  The images, model and residual for the best-fit lensing model are shown in Figure 2.
\newline\indent
With four images of the source, the model is well-constrained and the best-fit parameters are R$_{Ein}$ = 3.94$^{\prime\prime}$$^{+0.15}_{-0.11}$, R$_{core}$ = 0.5$^{\prime\prime}$$^{+0.4}_{-0.5}$, $e$ = 0.075$^{+0.015}_{-0.010}$, PA = -31.0$^{+0.3}_{-0.5}$ degrees, X$_{s}$ = 0.15$^{\prime\prime}$$^{+0.02}_{-0.01}$, and Y$_{s}$ = -0.11$^{\prime\prime}$$^{+0.01}_{-0.01}$.  The error bars have been calculated using 100 Monte Carlo simulations where the background noise in the image was resampled and the lensing model was re-fit.  The lensing model implies that the magnifications for the source images A, B, C, and D are factors of 4.5$^{+0.4}_{-0.6}$, 5.3$^{+0.7}_{-0.6}$, 4.6$^{+0.3}_{-0.5}$, and 2.1$^{+0.4}_{-0.3}$, respectively.
\subsection{The Size and Sersic Index of the Source Galaxy}
With the parameters of the lensing model constrained by the positions and brightnesses of the multiple images, we fix those parameters and re-perform the ray tracing and $\chi^2$-fitting of the ACS images.  This time allowing the R$_{e}$ and $n$ of the source galaxy as free parameters in the fit.  This produced a best-fit of R$_{e}$ = 0.43$^{+2.40}_{-0.40}$ kpc and n = 1.9$^{+2.1}_{-0.9}$.  Although formally fit as an ultra-compact galaxy with a profile intermediate between a disk and bulge, the uncertainties are extremely large due to the low S/N of the galaxy in the ACS F814W image.
\newline\indent
The S/N of the ground-based K$_{s}$-band image is substantially higher than the ACS image (see $\S$4) so we attempted to obtain better constraints on the size and profile using those data.  With the higher S/N data K$_{s}$-band data we measure an R$_{e}$ = 0.64$^{+0.08}_{-0.18}$ kpc and n = 2.2$^{+2.3}_{-0.9}$.  This is remarkably consistent with the ACS measurement despite the much larger uncertainties on the ACS-measured R$_{e}$.
\newline\indent
If we use the best-fit stellar mass of the system ($\S$ 5.2) then the galaxy lies a factor of $\sim$ 4.5 below the R$_{e}$ vs. stellar mass relation for galaxies in the local universe defined in \cite{vandokkum2008}.  If we compare to the stellar mass-size relation at $z \sim$ 2 defined in \cite{Bruce2012}, we find that it is consistent with the most compact galaxies in that sample.  Taken together this shows that the source galaxy appears to be a member of the massive ultra-compact galaxy population seen at $z \sim$ 2.   
\newline\indent
Unfortunately, at ground-based resolution (or with the current S/N in the ACS image) we cannot make stronger statements about the Sersic profile of the galaxy.  We note that while uncertain, the best-fit Sersic index, n = 2.2$^{+2.3}_{-0.9}$, suggesting that the galaxy that is intermediate between a disk and a bulge.  This value is also similar to the median $n$ of galaxies in the \cite{vandokkum2008} sample ($n$ = 2.3); however, we note that that sample has a large range of Sersic indexes (n = 0.5 - 4.5).  Unfortunately, better constraints on the light profile of the source will not be possible until higher S/N imaging with HST resolution is available.
\begin{deluxetable*}{lrrrrrc}
\tabletypesize{\footnotesize}
\scriptsize
\tablecaption{Photometric Data}
\tablewidth{7.0in}
\tablehead{\colhead{Filter} & \colhead{ Lens } & \colhead{ Source A } &  \colhead{ Source B } &
\colhead{Source C} & \colhead{Source D} & \colhead{Ap Corr}
\\
\colhead{} & \colhead{ Flux Density } & \colhead{ Flux Density} &  \colhead{ Flux Density} &
\colhead{Flux Density} & \colhead{Flux Density} & \colhead{}
\\
\colhead{(1)}& \colhead{(2)}& \colhead{(3)}& \colhead{(4)}&
\colhead{(5)}& \colhead{(6)}& \colhead{(7)}
}
\startdata
        u$^{*}$ &   1.28 $\pm$  0.47 &   0.06 $\pm$  0.84 &   0.35 $\pm$  0.84 &   0.37 $\pm$  0.84 &   0.29 $\pm$  0.84   &   2.33  \nl
            $B$ &   2.34 $\pm$  0.25 &   0.28 $\pm$  0.46 &   0.28 $\pm$  0.47 &   0.43 $\pm$  0.46 &   0.10 $\pm$  0.47   &   1.79  \nl
   $g^{\prime}$ &   3.12 $\pm$  0.30 &   0.49 $\pm$  1.14 &   0.64 $\pm$  1.15 &   1.52 $\pm$  1.14 &   0.01 $\pm$  1.15   &   3.33  \nl
            $V$ &   5.32 $\pm$  0.26 &   0.07 $\pm$  0.48 &   0.35 $\pm$  0.49 &   0.69 $\pm$  0.48 &   0.12 $\pm$  0.49   &   2.00  \nl
   $r^{\prime}$ &  10.82 $\pm$  0.29 &   0.57 $\pm$  0.43 &   0.71 $\pm$  0.44 &   0.88 $\pm$  0.43 &   0.36 $\pm$  0.45   &   2.08  \nl
   $i^{\prime}$ &  24.66 $\pm$  0.28 &   0.92 $\pm$  0.25 &   1.15 $\pm$  0.26 &   1.04 $\pm$  0.25 &   0.02 $\pm$  0.28   &   2.00  \nl
   $z^{\prime}$ &  41.25 $\pm$  0.39 &   0.53 $\pm$  0.64 &   0.91 $\pm$  0.66 &   1.48 $\pm$  0.64 &   0.56 $\pm$  0.68   &   2.33  \nl
        $IA427$ &   2.55 $\pm$  0.89 &   0.63 $\pm$  1.14 &   0.72 $\pm$  1.14 &  -0.27 $\pm$  1.14 &  -0.37 $\pm$  1.14   &   2.04  \nl
        $IA464$ &   1.75 $\pm$  0.77 &   0.93 $\pm$  1.81 &   0.97 $\pm$  1.81 &   0.21 $\pm$  1.81 &   0.96 $\pm$  1.81   &   2.94  \nl
        $IA484$ &   2.24 $\pm$  0.62 &   0.07 $\pm$  0.82 &   0.33 $\pm$  0.83 &   0.75 $\pm$  0.82 &   0.01 $\pm$  0.83   &   1.96  \nl
        $IA527$ &   4.50 $\pm$  0.68 &   0.02 $\pm$  0.67 &   0.49 $\pm$  0.68 &   0.59 $\pm$  0.67 &   0.00 $\pm$  0.68   &   1.89  \nl
        $IA505$ &   2.83 $\pm$  0.73 &  -0.66 $\pm$  0.99 &   0.31 $\pm$  0.99 &   0.23 $\pm$  0.99 &   0.52 $\pm$  0.99   &   2.04  \nl
        $IA574$ &   6.43 $\pm$  0.79 &   0.48 $\pm$  1.09 &   1.25 $\pm$  1.10 &   0.30 $\pm$  1.09 &   0.46 $\pm$  1.10   &   2.63  \nl
        $IA624$ &  10.17 $\pm$  0.74 &   0.49 $\pm$  0.57 &   0.40 $\pm$  0.57 &   0.80 $\pm$  0.57 &   0.06 $\pm$  0.59   &   1.89  \nl
        $IA679$ &  14.30 $\pm$  0.65 &   1.13 $\pm$  0.76 &   0.68 $\pm$  0.78 &   0.75 $\pm$  0.76 &   0.04 $\pm$  0.79   &   2.63  \nl
        $IA709$ &  16.08 $\pm$  0.76 &   1.11 $\pm$  0.59 &   1.04 $\pm$  0.59 &   1.30 $\pm$  0.58 &   0.83 $\pm$  0.61   &   2.38  \nl
        $IA738$ &  17.19 $\pm$  0.79 &   1.13 $\pm$  0.59 &   0.64 $\pm$  0.60 &   1.16 $\pm$  0.58 &   0.02 $\pm$  0.62   &   2.04  \nl
        $IA767$ &  18.22 $\pm$  0.77 &   0.58 $\pm$  0.86 &   0.57 $\pm$  0.87 &   1.58 $\pm$  0.86 &   0.26 $\pm$  0.89   &   2.94  \nl
        $IA827$ &  31.68 $\pm$  0.97 &   0.04 $\pm$  0.90 &   1.32 $\pm$  0.92 &   1.42 $\pm$  0.90 &  -0.03 $\pm$  0.96   &   3.23  \nl
            $Y$ &  44.47 $\pm$  0.48 &   1.24 $\pm$  1.10 &   2.44 $\pm$  1.13 &   2.05 $\pm$  1.10 &   0.83 $\pm$  1.17   &   2.44  \nl
            $J$ &  46.44 $\pm$  0.35 &   2.55 $\pm$  0.78 &   3.18 $\pm$  0.80 &   2.98 $\pm$  0.78 &   0.77 $\pm$  0.85   &   2.27  \nl
            $H$ &  44.78 $\pm$  0.28 &  11.06 $\pm$  0.57 &  12.00 $\pm$  0.58 &   9.88 $\pm$  0.57 &   5.65 $\pm$  0.61   &   2.13  \nl
        $K_{s}$ &  36.38 $\pm$  0.23 &  10.68 $\pm$  0.31 &  10.95 $\pm$  0.32 &  10.23 $\pm$  0.31 &   4.72 $\pm$  0.34   &   2.04  \nl
   $3.6\micron$ &  17.58 $\pm$  0.07 &   7.45 $\pm$  0.27 &   7.38 $\pm$  0.32 &   7.07 $\pm$  0.24 &   2.00 $\pm$  0.32   &   5.88  \nl
   $4.5\micron$ &   9.85 $\pm$  0.06 &   6.11 $\pm$  0.20 &   6.52 $\pm$  0.21 &   5.75 $\pm$  0.18 &   3.52 $\pm$  0.23   &   5.88  \nl
   $5.8\micron$ &   5.13 $\pm$  0.19 &   3.28 $\pm$  0.42 &   4.06 $\pm$  0.42 &   3.47 $\pm$  0.42 &   1.27 $\pm$  0.43   &   6.67  \nl
   $8.0\micron$ &   1.94 $\pm$  0.11 &   1.05 $\pm$  0.21 &   1.32 $\pm$  0.21 &   0.90 $\pm$  0.21 &   0.52 $\pm$  0.21   &   5.88  \nl
\enddata
\tablecomments{All flux densities are listed in units of 10$^{-19}$ ergs s$^{-1}$ cm$^{-2}$ \AA$^{-1}$.}
\end{deluxetable*}
\section{Photometric Data}
\subsection{Deblending Method}
\indent
The lens galaxy is bright and resolved at ground-based resolution.  Examination of the PSF-matched images shows that its extended light distribution causes non-negligible contamination to the photometry of the source galaxies.  In order to remove this contamination we used the GALFIT package \citep{CPeng2010} to model the lens/source system and subtract the flux of the lens.  For this procedure we do not use the PSF-matched images, which are convolved to the worst seeing image (the g$^{+}$ image, $\sim$ 1.2$^{\prime\prime}$ seeing).  Instead we use the original images, which have typical seeing of $<$ 1.0$^{\prime\prime}$ in the optical bands and $\sim$ 0.8$^{\prime\prime}$ in the NIR band. This subtraction procedure is illustrated in Figure 3, and is as follows.
\newline\indent
In the J, H, and K$_{s}$ bands the source galaxies A, B, C, and D are clearly detected and can be distinguished from the lens in the images (see Figures 1 and 3).  For these bands we used GALFIT to perform a simultaneous fit to all five galaxies.  Given the complexity of a five-object simultaneous fit, the process was somewhat unstable; however, we continuously refined the initial guess parameters until a good fit with null residuals for all five galaxies was achieved.  In these three bands the final fit values for the Sersic index ($n_{lens}$), effective radius (R$_{e, lens}$), axis ratio ($q_{lens}$), and position angle (PA$_{lens}$) of the central lens converged to similar values.  
\newline\indent
Once a good fit was determined, we subtracted only the model for the lens from the images.  The model for the lens galaxy and the residuals after subtraction in the K$_{s}$ and H bands are shown in Figure 3.  
\newline\indent
In the remaining 19 optical filters as well as the UltraVISTA Y-band the source galaxies are barely detected (see Figure 3) and attempts at a simultaneous five object fit using a wide range of initial guess parameters with GALFIT did not return reasonable values.  
In these bands we performed only a fit to the lens.  This fitting was stable and in all bands converged to similar values of $n_{lens}$, R$_{e, lens}$, $q_{lens}$, and PA$_{lens}$.  For these bands we subtracted the GALFIT model for the central lens (the Y-band subtraction is shown in Figure 3).  
\subsection{Optical and NIR Photometry}
With the contribution from the lens subtracted we performed aperture photometry on the residual images using SExtractor in dual image mode.  The source galaxies are brightest in the K$_{s}$-band so the K$_{s}$-band residual image was used as the detection image.  Photometry was performed in 1$^{\prime\prime}$ diameter apertures on the original seeing images in each band.  These images have different PSFs; however, using the PSF-matched images for photometry would be sub-optimal because they have had the good seeing data smoothed to the larger seeing disk of the worse seeing image.  Photometry on these images requires larger apertures that are more susceptible to contamination from residual flux missed in the fitting of the lens, as well as flux from the other source images which are separated by $\sim$ 2$^{\prime\prime}$.  We correct the photometry in each individual band to a total flux by assuming the source galaxies are point sources and extrapolating the growth curves of bright stars in the field.  Formally this is not completely correct because the source galaxy is resolved at ground-based resolution; however, as discussed in $\S$ 3, the R$_{e}$ of the source is a factor of two smaller than the PSF size, so a point-source provides a reasonable first-order approximation to the observed profile.
\newline\indent
The PSFs in the UltraVISTA YJHK$_{s}$ are remarkably similar and the seeing is 0.75$^{\prime\prime}$, 0.76$^{\prime\prime}$, 0.79$^{\prime\prime}$ and 0.82$^{\prime\prime}$ in the K$_{s}$, H, J, and Y bands, respectively \citep{McCracken2012}.  The correction to total magnitudes in these bands ranges between a factor of 2.04 and 2.44, hence the NIR colors do not depend strongly on the aperture corrections.  The PSF and seeing range in the optical is larger, 0.5$^{\prime\prime}$ -- 1.2$^{\prime\prime}$ \citep{Capak2007}.  The aperture corrections range between a factor of 1.79 and 3.33.  This is larger; however, as we discuss below, the source galaxies are not detected at $>$ 2$\sigma$ in the vast majority of the optical bands, (the exception is the deep COSMOS i$^{\prime}$ filter where they are detected at $\sim$ 4-5$\sigma$), so these bands provide only weak constraints on the SED of the galaxies.
\newline\indent
Photometry and associated errors for the source galaxies as well as the lens are listed in Table 1.  Calculation of the photometric errors includes three sources of uncertainty.  We estimate the uncertainty in the background subtraction by calculating the variance of the background in 1$^{\prime\prime}$ empty apertures in a regions around the lens/source system.  We also estimate a contribution from photon noise using the gain of the detector and the original exposure times, although we note this contribution is much smaller than the background variations for sources this faint.  Lastly, we perform aperture photometry on the GALFIT model at the location of each source.  We use the square root of this flux as an estimate in the additional uncertainty from subtraction of the model.  These three sources of uncertainty are added in quadrature to determine the total photometric error.  The aperture corrected photometry, the photometric errors, and the aperture corrections are listed in Table 1.
\subsection{IRAC Photometry}
The S-COSMOS IRAC imaging of the source is also deblended using GALFIT.  The source galaxies are clearly detected in all four IRAC channels; however, the large FHWM of the IRAC PSF creates stronger blends between the lens and source than in the optical and NIR bands and simultaneous five-object fits were highly unstable.  The problem was further complicated by the fact that the source galaxies are almost as bright as the lens in the 5.8$\micron$ and 8.0$\micron$ channels.  Therefore, instead of a simultaneous five object fit, we used the same approach as in the optical bands and fit only the central lens.  Given the strong blending we left the only free parameter in the fit as the flux of the lens, and constrained the position, $n_{lens}$, R$_{e, lens}$, $q_{lens}$, and PA$_{lens}$ using the best-fit GALFIT parameters from the NIR fits.  This constrained fit was much more stable and the residuals at the position of the lens after subtraction of the model were low.  
\newline\indent
With the lens subtracted aperture photometry was also performed in 1.0$^{\prime\prime}$ diameter apertures and corrected to a total flux using the growth curve of bright stars.  The aperture corrections for the IRAC photometry are substantially larger than the optical data, by a factor of $\sim$ 6 (see Table 1).  Although this correction is large, the IRAC PSF is well-determined and quite stable so the correction should be well-determined.  Regardless, we note that the inclusion of the IRAC photometry does not change the best-fit values of the parameters derived from the SED fitting ($\S$ 5.2).  It does considerably reduce the uncertainties in these parameters, particularly on the dust extinction.  This is typical for high-redshift massive galaxies \citep[e.g.,][]{Muzzin2009c}, and therefore we note that any uncertainties in the IRAC photometry will not change the interpretation of the source galaxy's SEDs.
\begin{figure*}
\plotone{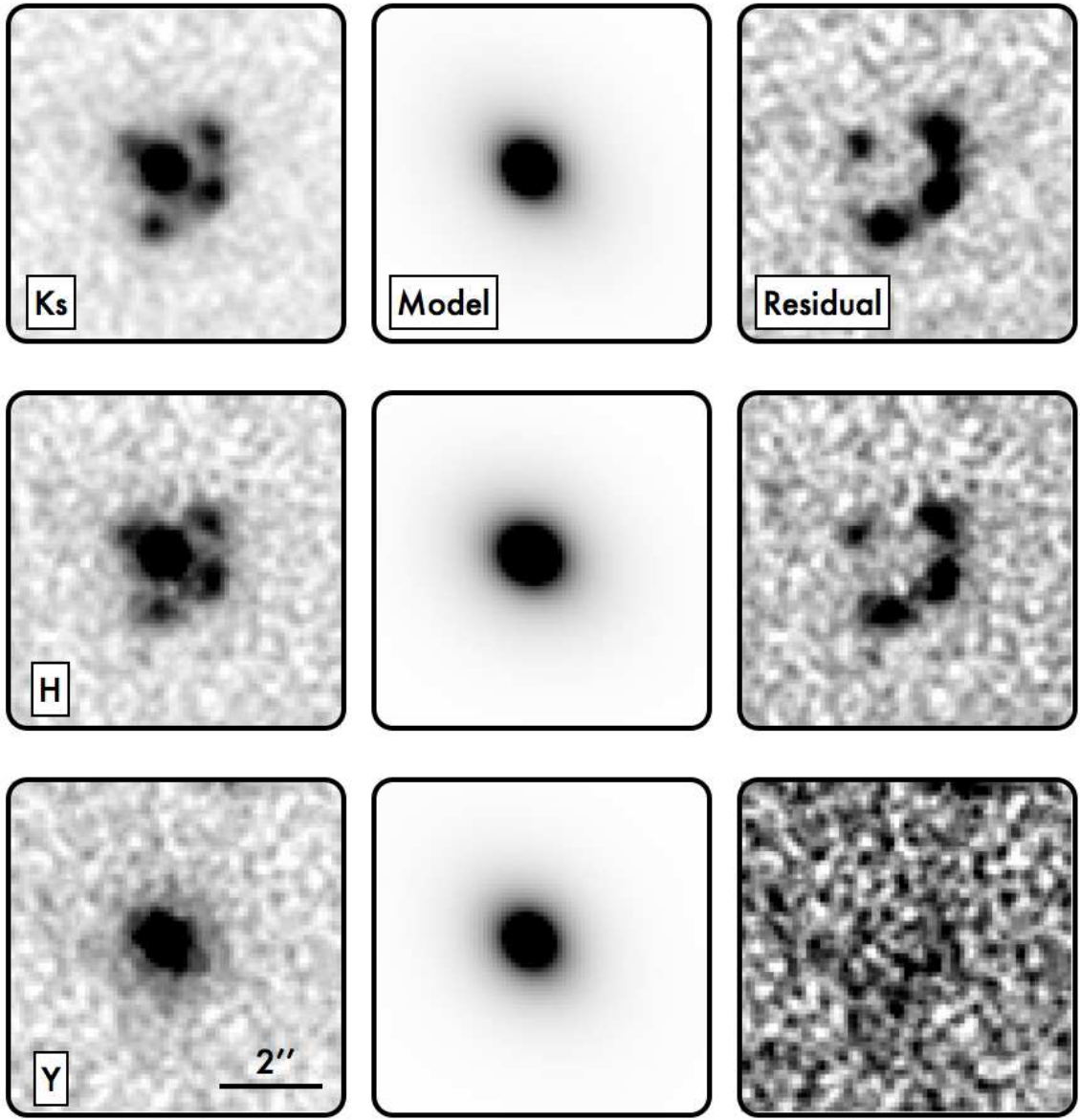}
\caption{\footnotesize Left Panels: Images of the lens/source system in the K$_{s}$, H, and Y bands.  Middle Panels: Models for the lens determined using GALFIT.  Right Panels: Residuals of subtracting the model for the lens.  Aperture photometry is performed on these residual images.}
\end{figure*}
\begin{figure*}
\plotone{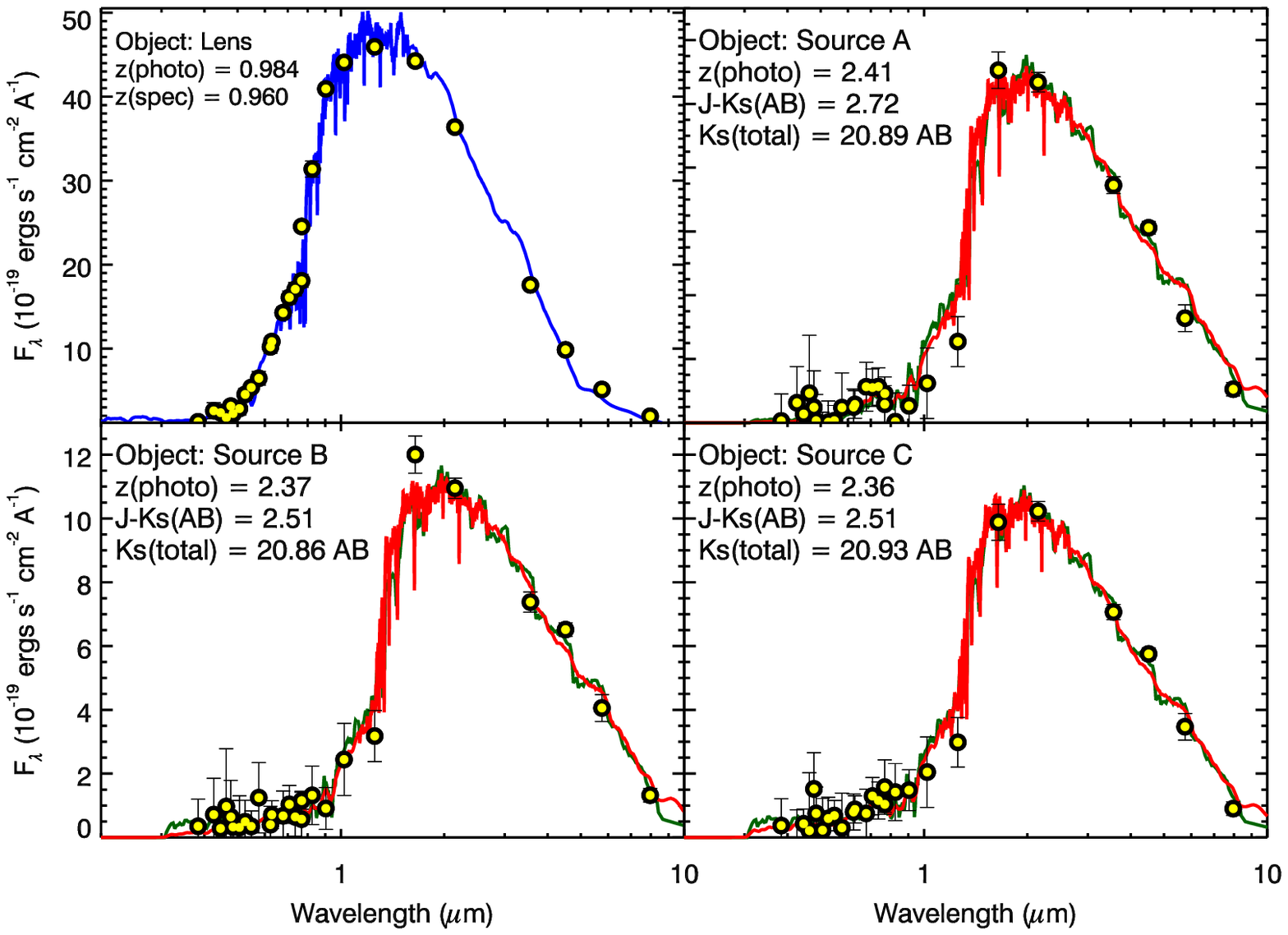}
\caption{\footnotesize Spectral energy distributions for the lens (upper left) and the three brightest images of the source galaxy (sources A, B, C).  The SEDs of the sources have been measured from aperture photometry on the images after subtraction of the central source using GALFIT.  The best-fit Bruzual \& Charlot model is plotted in blue for the lens, and in red for the source galaxies.  The best-fit Maraston model for the sources is shown in green.  The photometric redshifts of all three images are consistent at $z \sim$ 2.4.  The source galaxies are extremely red (J - K$_{s}$(AB) = 2.6), and are best-fit by (relatively) old templates with a  moderate amount of dust (A$_{v}$ $\sim$ 0.8). }
\end{figure*}
\section{Properties of the Lensed Galaxy}
\subsection{Photometric Redshifts}
\indent
Photometric redshifts ($z_{photo}$) for both the lens and source galaxy images are determined using the EAZY photometric redshift code \citep{Brammer2008}.  All twenty-seven photometric bands are used in the determination of the redshift, and the values are listed in Table 2.  The photometry for the source galaxies is derived from the deblended images, whereas we have derived the photometry for the lens from the original PSF-matched K$_{s}$-selected catalog.  The $z_{photo}$ of the lens galaxy is 0.98 $\pm$ 0.03 which agrees well with the spectroscopic redshift of $z = 0.960$ measured by \cite{Faure2011}.  The photometric redshifts for the images A, B, C, and D are 2.41 $\pm$ 0.13, 2.37 $\pm$ 0.18, 2.36 $\pm$ 0.15, and 2.22 $\pm$ 0.22, respectively, and are all consistent within the 1$\sigma$ uncertainties.  
\subsection{Stellar Mass, Age and Dust Content}
We perform SED fitting on the photometric data to determine the stellar population parameters using the FAST fitting code \citep{Kriek2009}.  For the SED fitting we assume the redshift as the best-fit $z_{photo}$ from EAZY.  We use the stellar population models from \cite{Bruzual2003} as our default model, but also fit to the \cite{Maraston2005} models.  We assume solar metallicity, a \cite{Calzetti2000} dust law, and a \cite{Chabrier2003} initial mass function.  We parameterize the star formation history as a declining exponential with an e-folding timescale of $\tau$.  Overall we fit four parameters, the stellar mass (M$_{star}$), age since the onset of star formation ($t$), $\tau$, and the V-band dust attenuation (A$_{v}$).  Integration of the star formation history also returns a star formation rate (SFR) and a star formation rate per unit stellar mass (specific star formation rate, SSFR).  The best-fit SEDs for both the lens and the three brightest sources: A, B, and C are plotted along with the corresponding photometry in Figure 4, and the best-fit stellar population parameters and associated errors are listed in Table 2.
\newline\indent
The lens is best-fit as an extremely massive (Log(M$_{star}$/M$_{\odot}$) = 11.49$^{+0.11}_{-0.19}$), old (Log(Age/yr$^{-1}$) = 9.4$^{+0.3}_{-0.3}$), and quiescent galaxy (Log(SSFR/yr$^{-1}$) = -11.70$^{+0.09}_{-99.9}$).  The M$_{star}$ derived from our photometry agrees reasonably well with the value of Log(M$_{star}$/M$_{\odot}$) = 11.64 $\pm$ 0.03 measured by \cite{Faure2011}.
\newline\indent
As Figure 4 shows, the SEDs of images A, B, and C are all similar within the uncertainties.  This is not necessarily expected, as small perturbations in the lens potential, and/or a non-symmetric source galaxy can results in different locations of the source being more or less magnified in each image \citep[e.g.,][]{Seitz1998,Sharon2012}.  We evaluate the overall best-fit stellar population parameters for the source by considering those derived from images A, B, C, which are the most strongly-magnified images. We take the median value of all parameters, but consider the full range of uncertainty allowed by the fits.  Source D, which is less magnified, is excluded in this calculation because the uncertainties are much larger than the other three images of the source.  We find that the source is massive (Log(M$_{star}$/M$_{\odot}$ = 10.8$^{+0.1}_{-0.1}$, corrected for lensing), old (Log(Age/yr$^{-1}$) = 9.0$^{+1.0}_{-0.4}$), and quiescent (Log(SSFR/yr$^{-1}$) = -11.41$^{+0.69}_{-99.9}$) with moderate dust extinction (A$_{v}$ = 0.8$^{+0.5}_{-0.6}$).  
\newline\indent
These stellar population parameters are quite typical of the well-studied population of massive ultra-compact quiescent galaxies at $z \sim$ 2 \citep[e.g.,][]{Kriek2008,vandokkum2008,Muzzin2009c,Muzzin2009a,vandesande2011,Cassata2011}. This suggests that the source is a likely to be a strongly-lensed member of that population.  The implied SSFR is extremely low; however, it is derived from the rest-frame UV/optical SED.  In order to test for any potential obscured star formation we examined the S-COSMOS MIPS 24$\micron$ image of the lens/source system.
\subsection{MIPS 24$\micron$ Photometry}
The S-COSMOS 24$\micron$ imaging is deep enough to detect galaxies with $\sim$ 100 M$_{\odot}$ yr$^{-1}$ of star formation at $z \sim$ 2.4.  Unfortunately, the FWHM resolution of the data are 5.5$^{\prime\prime}$, which is larger than the separation between both the source and lens, and the source galaxies themselves.  Given that individual sources cannot be resolved at 24$\micron$ we performed photometry in a 7$^{\prime\prime}$ diameter aperture to get a total flux density for the complete lens/source system.  Within this aperture we measure a flux of 26.3 $\pm$ 6.3 $\mu$Jy.  Based on the aperture corrections in the MIPS data handbook we correct this to a total flux of 67.3 $\pm$ 16.1 $\mu$Jy.
\newline\indent
The aperture photometry implies there is a significant 4$\sigma$ detection at the location of the lens/source system.  Examination of the image shows evidence for a faint object within the aperture; however, we note it is offset by a few arcseconds to the southwest of the lens/source system.  This clear offset suggests that the lens/source is not the correct counterpart.  There are however, no other plausible counterparts in the optical or NIR images near the 24$\micron$ detection, so if the lens/source is not the counterpart then the true counterpart must be highly extincted.
\newline\indent
We investigate the implications if the lens/source is the counterpart of the MIPS source.  If we assume that 100\% of the detected flux comes from the source and none comes from the lens, then the implied total SFR rate derived from the logarithmic average of the \cite{Dale2002} templates \cite[see][]{Wuyts2008,Muzzin2010}, uncorrected for lensing, is 111$^{+26}_{-27}$ $M_{\odot}$ yr$^{-1}$.  The total M$_{star}$ in all four images, uncorrected for lensing is log(M$_{star}/M_{\odot}$) = 11.97$^{+0.17}_{-0.07}$.  This implies a hard upper limit for the SSFR of the source as Log(SSFR) = -9.93$^{+0.20}_{-0.20}$.  This is $\sim$ 1.4 dex higher than the SSFR determined from the SED fitting (see Table 2).  If 100\% of the flux actually comes from the source, then it implies that there may be some optically thick star formation continuing within the galaxy, or that it contains an obscured AGN, or both.  We also checked if the lens/source was detected in either the $XMM$ X-ray and VLA radio observations and catalogs of the COSMOS field but found no detection at the location of the lens/source.  
\newline\indent
Recent papers have shown evidence for a ``main sequence" of star forming galaxies out to $z \sim$ 3 \citep[e.g.,][]{Noeske2007,Elbaz2007,Wuyts2011,Whitaker2012}.   We note that while the MIPS flux of the source galaxies may imply it is not completely quiescent, the typical SSFR of galaxies with Log(M$_{star}$/M$_{\odot}$) = 10.8 at $z \sim$ 2.5 is Log(SSFR) = -8.6 yr$^{-1}$ \citep{Whitaker2012}.  This is more than an order of magnitude higher than the upper limits on the SSFR, suggesting that if the source is a star forming galaxy, it lies well off the star forming main sequence and may be en route to becoming fully quenched. 
\newline\indent
We also cannot rule out the possibility that some or all of the 24$\micron$ flux comes from the lens.  The SED-measured SFR of the lens is only 0.6 M$_{\odot}$ yr$^{-1}$, which would only account for 4 $\mu$Jy, or 6\% of the observed flux.  However, if we note that only 10 M$_{\odot}$ yr$^{-1}$ of obscured star formation in the lens could account for 100\% of the observed 24$\micron$ flux.  Likewise, an obscured AGN could account for a fraction or all of the total flux.
\section{Discussion and Conclusion}
\begin{deluxetable*}{lccccc}
\tabletypesize{\footnotesize}
\scriptsize
\tablecaption{Stellar Population Parameters}
\tablewidth{7.0in}
\tablehead{\colhead{Parameter} & \colhead{ Lens } & \colhead{ Source A } &  \colhead{ Source B } &
\colhead{Source C} & \colhead{Source D}
\\
\colhead{(1)}& \colhead{(2)}& \colhead{(3)}& \colhead{(4)}&
\colhead{(5)}& \colhead{(6)}}
\startdata
\multicolumn{6}{c}{Bruzual \& Charlot 2003 Models}\\
\hline
$z_{photo}$ & 0.98 $\pm$ 0.03 & 2.41 $\pm$ 0.13 & 2.37 $\pm$ 0.18 & 2.36 $\pm$ 0.15 & 2.22 $\pm$ 0.22 \nl
Log(Stellar Mass) (M/M$_{\odot}$) & 11.49$^{+0.11}_{-0.19}$ & 10.82$^{+0.05}_{-0.07}$(11.48) & 10.78$^{+0.07}_{-0.10}$(11.50) & 10.70$^{+0.17}_{-0.01}$(11.36) & 10.68$^{+0.16}_{-0.17}$(11.00) \nl
Log($\tau$) (yr) & 8.5$^{+0.3}_{-1.5}$ & 8.0$^{+0.3}_{-1.0}$ & 8.1$^{+0.5}_{-1.1}$ & 8.2$^{+0.3}_{-0.3}$ & 7.0$^{+1.6}_{-0.0}$ \nl
Log(Age) (yr) & 9.4$^{+0.3}_{-0.3}$ & 9.0$^{+0.2}_{-0.2}$ & 9.0$^{+0.4}_{-0.2}$ & 9.1$^{+0.4}_{-0.2}$ & 9.0$^{+0.5}_{-0.2}$ \nl
A$_{v}$ & 0.6$^{+0.3}_{-0.4}$ & 0.9$^{+0.2}_{-0.6}$ & 1.0$^{+0.2}_{-0.7}$ & 0.6$^{+0.5}_{-0.4}$ & 0.8$^{+0.5}_{-0.8}$ \nl
Log(SFR) (M$_{\odot}$ yr$^{-1}$) & -0.21$^{+0.07}_{-99.9}$ & -1.26$^{+0.81}_{-99.9}$(-0.61) & -0.54$^{+0.13}_{-99.9}$(0.18) & -0.71$^{+0.82}_{-0.07}$(-0.05) & -99.9$^{+99.5}_{-0.0}$(-99.9) \nl
Log(SSFR) (yr$^{-1}$) & -11.70$^{+0.09}_{-99.9}$ & -12.10$^{+0.82}_{-99.9}$ & -11.32$^{+0.07}_{-99.9}$ & -11.41$^{+0.69}_{-0.20}$ & -99.9$^{+87.60}_{-0.0}$ \nl
\hline
\multicolumn{6}{c}{Maraston 2005 Models}\\
\hline
$z_{photo}$ & 0.98 $\pm$ 0.03 & 2.41 $\pm$ 0.13 & 2.37 $\pm$ 0.18 & 2.36 $\pm$ 0.15 & 2.22 $\pm$ 0.22 \nl
Log(Stellar Mass) (M/M$_{\odot}$) & 11.40$^{+0.00}_{-0.00}$ & 10.76$^{+0.17}_{-0.09}$(11.41) & 10.69$^{+0.16}_{-0.14}$(11.41) & 10.68$^{+0.13}_{-0.08}$(11.34) & 10.72$^{+0.16}_{-0.14}$(11.04) \nl
Log($\tau$) (yr) & 8.7$^{+0.0}_{-0.0}$ & 8.2$^{+0.3}_{-1.2}$ & 8.3$^{+0.2}_{-1.3}$ & 8.3$^{+0.2}_{-0.2}$ & 7.9$^{+0.7}_{-0.9}$ \nl
Log(Age) (yr) & 9.6$^{+0.0}_{-0.0}$ & 9.2$^{+0.2}_{-0.2}$ & 9.2$^{+0.4}_{-0.2}$ & 9.2$^{+0.4}_{-0.2}$ & 9.3$^{+0.2}_{-0.1}$ \nl
A$_{v}$ & 0.0$^{+0.0}_{-0.0}$ & 0.2$^{+0.2}_{-0.2}$ & 0.3$^{+0.2}_{-0.3}$ & 0.2$^{+0.2}_{-0.2}$ & 0.2$^{+0.4}_{-0.2}$ \nl
Log(SFR) (M$_{\odot}$ yr$^{-1}$) & -0.53$^{+0.00}_{-0.00}$ & -0.97$^{+0.04}_{-27.5}$(-0.32) & -1.64$^{+0.81}_{-99.9}$(-0.92) & -0.79$^{+0.07}_{-99.9}$(-0.13) & -8.22$^{+7.1}_{-99.9}$(-7.56) \nl
Log(SSFR) (yr$^{-1}$) & -11.94$^{+0.00}_{-0.00}$ & -12.32$^{+0.77}_{-99.9}$ & -11.55$^{+0.10}_{-99.9}$ & -11.55$^{+0.10}_{-0.10}$ & -18.59$^{+7.10}_{-99.9}$ \nl
\enddata
\tablecomments{The values listed for Log(Stellar Mass) and Log(SFR) are corrected based on the lensing model.  The values in brackets are the direct measurements with no correction for lensing.}
\end{deluxetable*}

Overall, the size constraints from the lens modeling as well as the SED modeling confirm that the strongly-lensed galaxy COSMOS 0050+4901 is in fact the first example of a strongly-lensed massive ultra-compact quiescent galaxy.  The high magnification from the strong lensing implies that with future high S/N, high-resolution imaging we may now have the possibility to peer into the cores of these compact systems.  Likewise the brightening effects should allow us to determine a high-quality velocity dispersion with a reasonable integration time on an 8m-class telescope.
\newline\indent
Discovering a larger representative sample of strongly-lensed massive ultra-compact quiescent galaxies would be an important step forward to obtaining a more detailed picture of their evolution.  It would also allow comparison studies with the samples of strongly-lensed blue star-forming and submillimeter galaxies that already exist.  One question that needs to be answered in order to find such samples is how frequently these lensed sources appear, and therefore how much area with high-quality NIR imaging needs to be searched?  This is a difficult question to answer precisely, as it requires knowledge of the underlying set of lenses and sources, as well as an understanding of the observational selection effects.  Here we present an order-of-magnitude estimate based on what is currently known about current strong lensing samples, the stellar mass function, and the fact that we have discovered one such source in the UltraVISTA field.  
\newline\indent
If we make the optimistic assumption that searches for strong lenses can identify all strongly-lensed galaxies down to Log(M$_{star}$/M$_{\odot}$) = 9.5 \citep[these are the lowest-mass galaxies that have been identified with strong lensing, e.g.,][]{Rigby2011,Wuyts2012}, then the fraction of high-redshift strong lenses that will be massive galaxies at $z \sim$ 2 can be estimated using the stellar mass function.  If we integrate the stellar mass function at 2 $< z < $ 3 from \cite{Marchesini2009}, we find that the ratio of galaxies with log(M$_{star}$/$M_{\odot}$) $>$ 11.0 compared to those with 9.5 $<$ log(M$_{star}$/M$_{\odot}$) $<$ 11.0 is a factor of 20.7. This suggests that roughly one in twenty high-redshift strong lenses will be a massive galaxy, although the true number recovered will depend on observational selection effects.  Of these massive galaxies, only $\sim$ 50\% will be a quiescent \citep[e.g.,][]{Kriek2008,Muzzin2009c}. 
\newline\indent
In their ``high-quality" sample of 16 strong lenses in the COSMOS field from \cite{Faure2008} estimate that 9/16 strong lenses are at $z >$ 2, and this is consistent with the median redshift of the brightest strongly-lensed galaxies in cluster surveys \citep[e.g.,][]{Bayliss2011a,Bayliss2011b}.    
\newline\indent
To estimate the frequency of strongly-lensed massive quiescent galaxies at $z >$ 2, we use the total sample of 67 strong lenses in \cite{Faure2008} as an estimate of the total number of strongly-lensed galaxies per COSMOS field ($\sim$ 1.5 deg$^2$).  Using this empirically-derived number as a starting point does account for some of the observational selection effects.  The total number of strongly-lensed massive quiescent galaxies that are expected within the COSMOS field should be (total lenses) $\times$ (fraction of lenses at $z >$ 2) $\times$ (fraction of lensed sources that are log(M$_{star}$/M$_{\odot}$) $>$ 11.0) $\times$ (fraction of log(M$_{star}$/M$_{\odot}$) $>$ 11.0 galaxies that are quiescent).  Filling in the numbers above gives 67 $\times$ 0.56 $\times$ 0.05 $\times$ 0.5 = 0.94.  The number is of order unity suggesting there should be approximately one such source per COSMOS field, consistent with the single source we have discovered.  This agreement is comforting, and suggests that the order-of-magnitude calculation is reasonable; however, all of the numbers used in the estimate are highly uncertain, and most likely there are additional selection effects that are not accounted for.  
\newline\indent
The true frequency of strongly-lensed massive ultra-compact quiescent galaxies may be factors of several larger or smaller.  Still, the fact that we found only one source in the COSMOS/UltraVISTA survey would seem rule out the possibility that these sources are significantly more abundant than our estimate.  Also, it is worth noting that no other such sources have yet been reported in similar sized surveys such as the UDS \citep[]{Williams2009} and NMBS \citep[]{Whitaker2011}, which would again support the idea that the abundance is not much higher than this estimate.  
\newline\indent
It is possible that the abundance is lower than our estimate, and that it is fortuitous to find such a system in the COSMOS/UltraVISTA field.  The lack of such sources in the UDS, and in the 50\% of the NMBS that is not in COSMOS field means that the current total area of deep-wide NIR surveys is $\sim$ 2.5 deg$^2$, with only one lensed massive, ultra-compact, quiescent galaxy discovered so far.  It is difficult to say more without better statistics, but if the order-of-magnitude estimate is correct it suggests that the space density of strongly-lensed massive quiescent galaxies at $z >$ 2 is roughly one per 1-2 deg$^2$.  This suggests that in order to obtain a sample of 10 such galaxies, medium-deep optical/NIR photometry with good angular resolution (to avoid lens/source blending) covering $\sim$ 10 - 20 deg$^2$ would be required.  Optical/NIR surveys with these requirements are currently being performed, which suggests that if searches are careful, the prospects are good for obtaining a real sample of strongly-lensed massive ultra-compact quiescent galaxies within the next few years.
\acknowledgements
JSD acknowledges the support of the European Research Council via the award of an Advanced Grant, and the support of the Royal Society through a Wolfson Research Merit Award.  BMJ and JPUF acknowledge support from the ERC-StG grant EGGS-278202.
The Dark Cosmology Centre is funded by the Danish National Research
Foundation.

\bibliographystyle{apj}
\bibliography{apj-jour,myrefs}




\end{document}